\documentclass[preprint,superscriptaddress,floatfix,amssymb,amsmath]{revtex4}
\usepackage[dvips]{graphicx}
\usepackage{epsfig}
\usepackage{tabularx}
\usepackage{pifont}
\usepackage{color}
\usepackage[pdfpagemode=UseNone,colorlinks=true,linkcolor=blue,citecolor=blue]{hyperref}

\newcommand{\xzpf}{x_{\mathrm{zpf}}}
\newcommand{\bh}{\hat{b}}
\newcommand{\bd}{\hat{b}^\dagger}
\newcommand{\nm}{\bar{n}}
\newcommand{\nth}{\bar{n}_{\mathrm{th}}}

\newcommand{\Geff}{\Gamma_{\mathrm{eff}}}

\newcommand{\Gm}{\Gamma_{\mathrm{m}}}
\newcommand{\Gplus}{\Gamma_{\mathrm{+}}}
\newcommand{\Gminus}{\Gamma_{\mathrm{-}}}
\newcommand{\Rplus}{R_{+}}
\newcommand{\Rminus}{R_{-}}
\newcommand{\Omegam}{\Omega_{\mathrm{m}}}
\newcommand{\Dpump}{\Delta_{\mathrm{pump}}}
\newcommand{\epsc}{\epsilon_{\mathrm{c}}}

\newcommand{\Wpar}{\Omega_{\mathrm{par}}}

\begin{document}

\title{Quantum signature of a squeezed mechanical oscillator}

\author{A. Chowdhury} 
\affiliation{CNR-INO, L.go Enrico Fermi 6, I-50125 Firenze, Italy}

\author{P. Vezio}
\affiliation{European Laboratory for Non-Linear Spectroscopy (LENS), Via Carrara 1, I-50019 Sesto Fiorentino (FI), Italy}

\author{M. Bonaldi}
\affiliation{Institute of Materials for Electronics and Magnetism, Nanoscience-Trento-FBK Division,
 38123 Povo, Trento, Italy}
\affiliation{Istituto Nazionale di Fisica Nucleare (INFN), Trento Institute for Fundamental Physics and Application, I-38123 Povo, Trento, Italy}

\author{A. Borrielli}
\affiliation{Institute of Materials for Electronics and Magnetism, Nanoscience-Trento-FBK Division,
 38123 Povo, Trento, Italy}
\affiliation{Istituto Nazionale di Fisica Nucleare (INFN), Trento Institute for Fundamental Physics and Application, I-38123 Povo, Trento, Italy}

\author{F. Marino}
\affiliation{INFN, Sezione di Firenze}
\affiliation{CNR-INO, L.go Enrico Fermi 6, I-50125 Firenze, Italy}

\author{B. Morana}
\affiliation{Institute of Materials for Electronics and Magnetism, Nanoscience-Trento-FBK Division,
 38123 Povo, Trento, Italy}
\affiliation{Dept. of Microelectronics and Computer Engineering /ECTM/DIMES, Delft University of Technology, Feldmanweg 17, 2628 CT  Delft, The Netherlands}

\author{G. A. Prodi}
\affiliation{Istituto Nazionale di Fisica Nucleare (INFN), Trento Institute for Fundamental Physics and Application, I-38123 Povo, Trento, Italy}
\affiliation{Dipartimento di Fisica, Universit\`a di Trento, I-38123 Povo, Trento, Italy}

\author{P.M. Sarro}
\affiliation{Dept. of Microelectronics and Computer Engineering /ECTM/DIMES, Delft University of Technology, Feldmanweg 17, 2628 CT  Delft, The Netherlands}

\author{E. Serra}
\affiliation{Istituto Nazionale di Fisica Nucleare (INFN), Trento Institute for Fundamental Physics and Application, I-38123 Povo, Trento, Italy}
\affiliation{Dept. of Microelectronics and Computer Engineering /ECTM/DIMES, Delft University of Technology, Feldmanweg 17, 2628 CT  Delft, The Netherlands}

\author{F. Marin}
\email[Electronic mail: ]{marin@fi.infn.it}
\affiliation{CNR-INO, L.go Enrico Fermi 6, I-50125 Firenze, Italy}
\affiliation{European Laboratory for Non-Linear Spectroscopy (LENS), Via Carrara 1, I-50019 Sesto Fiorentino (FI), Italy}
\affiliation{INFN, Sezione di Firenze}
\affiliation{Dipartimento di Fisica e Astronomia, Universit\`a di Firenze, Via Sansone 1, I-50019 Sesto Fiorentino (FI), Italy}

\date{\today}
\begin{abstract}
Some predictions of quantum mechanics are in contrast with the macroscopic realm of everyday experience, in particular those originated by the Heisenberg uncertainty principle, encoded in the non-commutativity of some measurable operators.
Nonetheless, in the last decade opto-mechanical experiments have actualized  macroscopic mechanical oscillators exhibiting such non-classical properties. A key indicator is the asymmetry in the strength of the motional sidebands generated in an electromagnetic field that measures interferometrically the oscillator position.
This asymmetry is a footprint of the quantum motion of the oscillator, being originated by the non-commutativity between its ladder operators. 
A further step on the path highlighting the quantum physics of macroscopic systems is the realization of strongly non-classical states and the consequent observation of a distinct quantum behavior.
Here we extend indeed the analysis to a squeezed state of a macroscopic mechanical oscillator embedded in an optical cavity, produced by parametric effect originated by a suitable combination of optical fields. The motional sidebands assume a peculiar shape, related to the modified system dynamics, with asymmetric features revealing and quantifying the quantum component of the squeezed oscillator motion.
\end{abstract}

%\pacs{42.50.Wk,45.80.+r,05.40.-a,07.10.Cm}

\maketitle

The quantum description of the microscopic world is verified with ever increasing accuracy, and more and more common and indispensable technologies are based on intrinsically quantum principles. On the other hand, some features predicted by quantum mechanics are in contrast with the everyday experience, in particular those originated by entanglement and by the Heisenberg uncertainty principle, encoded in the non-commutativity of some measurable operators. It is therefore not only interesting from the scientific and technological point of view, but also useful for extending our intuitive grasp of quantum devices, to verify how such behaviors are conserved and/or modified in the transition between microscopic and macroscopic reality.

A relevant contribution in this direction is being given by the experimental investigation in the field of quantum opto-mechanics  \cite{AspelRMP}, that allowed in the last decade to realize systems of macroscopic mechanical oscillators exhibiting non-classical properties. A key example is the observation of an asymmetry in the strength of the motional sidebands generated in an electromagnetic field that measures interferometrically the position of the mechanical oscillator \cite{Safavi2012,Purdy2015,Underwood2015,Peterson2016,Sudhir2017a,Qiu2019}. 
%\viola{If the probe field interacts  weakly enough to avoid significantly modifying the state of the oscillator, and} 
As far as spurious experimental features are avoided \cite{Jayich2012,Safavi2013,Qiu2018}, this asymmetry becomes a footprint of the quantum motion of the oscillator, being originated by the non-commutativity between its ladder operators \cite{Khalili2012,Weinstein2014,Borkje2016}. The sideband asymmetry is measurable whenever the thermal agitation is weak enough, and therefore for low phonon number.
%, even if a somehow similar indicator (i.e., the correlation between phase and amplitude fluctuations of the probe) has shown in recent experiments a quantum signature up to room temperature \cite{Purdy2017,Sudhir2017}. 

A further step on the path highlighting quantum features in macroscopic systems is the realization of strongly non-classical states and the consequent evidence of specific quantum indicators. In this work we turn our attention to the squeezed state of a macroscopic mechanical oscillator embedded in an optical cavity. Squeezing is generated by the parametric effect originated by a suitable combination of optical fields. We show that the motional sidebands assume a particular shape, related to the modified system dynamics, and that their asymmetry, an essentially non-classical phenomenon, presents a clear signature of the quantum squeezing of the oscillator.

The coordinate expressing the position of an harmonic oscillator can be written as $x = 2 \xzpf \left[X \cos (\Omegam t + \mathrm{\phi})+ Y \sin (\Omegam t + \mathrm{\phi})\right]$ where $\xzpf$ is 
%a normalization length that, in view of a quantum description, can be taken as equal to 
the ground state spread of the position (i.e., $\xzpf = \sqrt{\hbar/2 m \Omegam}$ where $m$ is the mass and $\Omegam$ the frequency of the oscillator). The quadratures $X$ and $Y$ are slowly varying if the oscillator is damped. In a thermal state, the variance of the quadratures is $\langle X^2 \rangle  = \langle Y^2 \rangle  = (2\nth+1)/4$ where $\nth$ is the mean thermal occupation number, for any choice of the phase $\mathrm{\phi}$. The mechanical interaction of the oscillator with a readout electromagnetic field produces in the latter motional sidebands around its main field frequency, displaced by $\pm \Omegam$, and proportional respectively to $\bh = X + i Y$ (anti-Stokes sideband) and $\bd = X-iY$ (Stokes sideband). In a thermal state, the variances of $\bh$ and $\bd$ are respectively $\langle \bd \, \bh \rangle  = \nth$ and $\langle \bh \, \bd \rangle  = \nth +1$ \cite{Clerk2010}. In cavity opto-mechanics experiments, the electromagnetic field competes with the thermal environment yielding, in appropriate conditions, a larger overall coupling rate $\Geff$, a mechanical resonance frequency modified by the optical spring effect, and a lower mean phonon number $\nm$ \cite{AspelRMP,Arcizet2006}. 
%In the last decade, such optical cooling has been pushed to reduce $\nm$ around or even well below unity  \cite{oconnell10,teufel11,chan11,verhagen12,Purdy2012}. 
The ratio between Stokes and anti-Stokes sidebands can now be written as $R = (\nm+1)/\nm$ and, for low enough $\nm$,
a deviation from unity of $R$ becomes measurable, providing a clear signature of the smooth transition between the classical motion and the quantum behavior. In case of continuous measurements on a weakly coupled oscillator (with $\Geff  \ll  \Omegam$), the observed variables (either $x$, or the quadratures, or the sidebands) exhibit Lorentzian spectra with width $\Geff$, and their variance can be evaluated by integrating over the respective spectral peaks. 

In the absence of an intrinsic reference phase, all the quadratures (for any choice of $\mathrm{\phi}$) are obviously indistinguishable. On the other hand, a phase-sensitive interaction can break such symmetry and provide, e.g., different variances in two orthogonal quadratures. In this case the oscillator is said to be in a squeezed state. 
A scheme that allowed to obtain it in cavity opto-mechanical experiments is the so-called reservoir engineering \cite{Verstraete2009,Kronwald2013}. In this configuration, unbalanced fields are detuned respectively by $\pm \Omegam$ from the cavity resonance. Due to the interference between the noise sidebands of the two fields, the electromagnetic reservoir seen by the mechanical oscillator is modified in a phase-sensitive way. When its effect dominates over the thermal one, the result is a squeezed oscillator. It has been experimentally achieved in microwave experiments with cooled nano oscillators \cite{Wollman2015,Pirkkalainen2015,Lecocq2015,Lei2016}, where the variance in one quadrature was indeed reduced below that of the ground state (i.e., $\langle X^2 \rangle  < 1/4$ for a particular value of $\mathrm{\phi}$). This property was verified through accurate calibrations of the spectrum of the electromagnetic field that had interacted with the oscillator. This noise reduction can play an important role, e.g., in optimizing the sensitivity of the oscillator used as quantum sensor. However, we point out that the motional sidebands remain unmodified and a more direct evidence of a non-classical behavior is lacking.

A different possibility to produce a squeezed state of the oscillator is the modulation of its spring constant at twice its resonance frequency (parametric modulation) \cite{Rugar1991}. This scheme has been implemented in several experiments with thermal oscillators, including cavity opto-mechanical setups where a modulation of the optical spring is obtained by acting on the light intensity or frequency \cite{Pontin_PRL2014,Sonar2018}. The variances in two, suitably chosen, orthogonal quadratures are altered, respectively, by factors of $1/(1+s)$ and $1/(1-s)$, where $s$ is the parametric gain. The correspondent spectra keep Lorentzian shapes with modified widths, given respectively by $\Gplus = \Geff (1+s)$ (over-damped quadrature) and $\Gminus = \Geff (1-s)$ (under-damped quadrature). 
It is thus interesting to consider what happens to the spectra of the motional sidebands, that are somehow given by linear combinations of the two quadratures. In particular, we remark that the sidebands provide a quantum signature of the oscillator motion even when the thermal effect is still dominating (sideband asymmetry was measured even with $\nm \simeq 100$ \cite{Sudhir2017a}), therefore they represent a powerful indicator that one would hope to exploit to also show non-classical features of squeezed states. Not surprisingly, the spectrum of each sideband turns out to be composed of two Lorentzian curves, with width $\Gplus$ and $\Gminus$. More relevant, the areas of this two Lorentzian components are different in the two sidebands, with ratios given by
\begin{eqnarray}
\Rplus = \frac{\nm +1+s/2}{\nm -s/2}    \label{ratios1}\\
\Rminus = \frac{\nm +1-s/2}{\nm +s/2}   \label{ratios2}
\end{eqnarray}
respectively for the broader ($\Rplus$) and narrower ($\Rminus$) components. If the oscillator motion is described by classical (commuting) variables, the spectra corresponding to the two motional sidebands must be identical, and the same happens if we give a quantum description of a ``classical'' (i.e., thermal noise dominated, with $\nm  \gg  1$) oscillator. On the other hand, for moderately low $\nm$ the sideband ratios $\Rplus$ and $\Rminus$ differ not only from unity, but also from the ratio $R$ measurable without parametric squeezing. Namely, the ratio is stronger for the broadened Lorentzian component, while it approaches the unity for the narrowed component as $s \to 1$ (i.e., close to the parametric instability threshold).
Therefore, a purely quantum effect can be put into evidence even for a state having a variance exceeding that of the ground state in any quadrature and, besides thermal noise, even for states that are not of minimal uncertainty (i.e., with $\langle X^2 \rangle \langle Y^2 \rangle  > 1/16$) as those created by parametric squeezing. 

In the following, we describe an experimental study of this effect, and we show that a non-classical state of the macroscopic mechanical oscillator is realized through interaction with optical fields.

\begin{figure}
\begin{centering}
\includegraphics[scale=0.5]{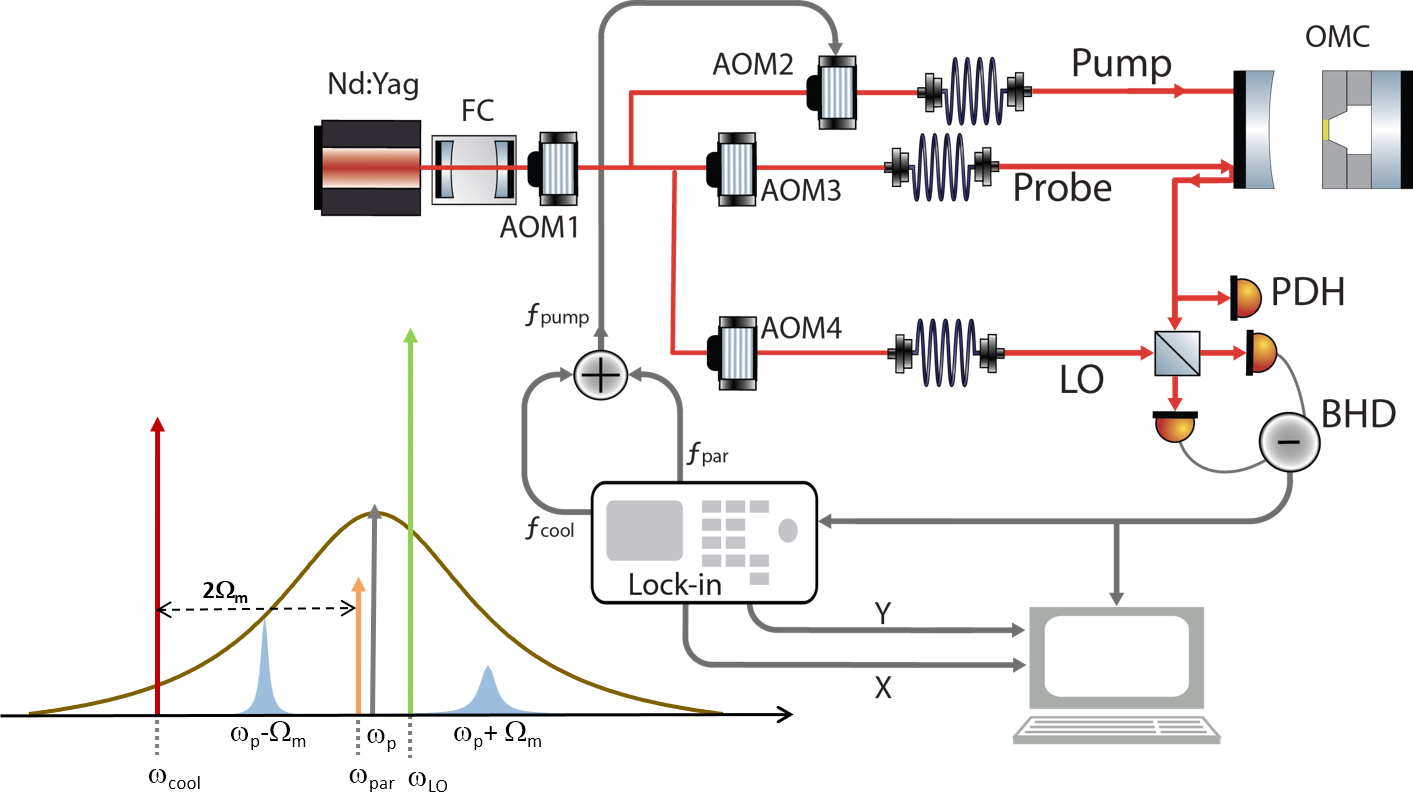}
%\par
\caption{Simplified scheme of the experimental setup. The light of a Nd:YAG laser is filtered by a cavity having a linewidth of $66\,$kHz, frequency tuned by a first acousto-optic modulator (AOM1), 
and split into three parts, frequency shifted by three further AOMs.
A first beam (probe) is phase modulated at $13\,$MHz for the 
Pounder-Drever-Hall (PDH) locking to the resonance of the optomechanical
cavity (OMC). The second beam is the local oscillator (LO) in the heterodyne detection of the reflected probe. Finally, the pump beam contains two frequencies, a stronger cooling tone (at $\omega_{\mathrm{cool}}$) and a parametric modulation tone (at $\omega_{\mathrm{par}}$). 
After single-mode fibers, pump and probe are combined with orthogonal polarizations and mode-matched to the OMC. About $2 \mu$W of the reflected probe are sent to the PDH detection, while most of the probe light is combined with the LO in a balanced detection (BHD). Also shown is a scheme of the field frequencies. The LO is placed on the blue side of
the probe ($\omega_{\mathrm{p}}$) and detuned by $\Delta_{\mathrm{LO}}  \ll  \Omegam$, therefore the Stokes lines are on the red
side of the LO, while the anti-Stokes lines are on the blue side. In the heterodyne spectra, they are located respectively at $\Omegam+\Delta_{\mathrm{LO}}$ (Stokes) and $\Omegam-\Delta_{\mathrm{LO}}$ (anti-Stokes).}
\label{fig_setup}
\end{centering} 
\end{figure}

The experimental setup is sketched in Figure (\ref{fig_setup}). The measurements are performed on a circular SiN membrane with a thickness
of $100\,$nm and a diameter of $1.64\,$mm, supported by a silicon ``loss shield'' structure \cite{Borrielli2014,Borrielli2016,Serra2016,Serra2018}. In this work we exploit the (0,2) mechanical mode at $\sim530$ kHz, having a quality factor of $6.4 \times 10^6$ at cryogenic temperature.  
 
The oscillator is
placed in a Fabry-Perot cavity of length $4.38$~mm, at $2$~mm from the cavity flat end mirror, forming a ``membrane-in-the-middle'' setup \cite{Jayich2008}. The input coupler is concave with a radius of 50~mm, originating a waist of 70~$\mu$m. The cavity finesse and linewidth are respectively $24500$ and $\kappa=1.4 \ \textrm{MHz} \times 2\pi$. The optomechanical cavity is cooled down to $\sim 7\,$K in an helium flux cryostat.

The light of a Nd:YAG laser is filtered by a Fabry-Perot cavity and split into three beams, whose frequencies are controlled by means of acousto-optic modulators (AOM) (Fig. \ref{fig_setup}a).
The first
beam (probe, at frequency $\omega_{\mathrm{p}}$) is always resonant with the optomechanical cavity. It is kept locked using the Pound-Drever-Hall detection and a servo loop that exploits the first AOM to follow fast fluctuations, and a piezo-electric transducer to compensate for long term drifts of the cavity length. 

A second beam is used as local oscillator (LO) in a balanced detection of the probe beam, reflected by the cavity. The LO frequency $\omega_{\mathrm{LO}}$ is shifted with respect to the probe
(namely, by $\Delta_{\mathrm{LO}}/2\pi = 11\,$kHz), to realize a low-frequency heterodyne detection \cite{Pontin2018a}. The output of the balanced detection is acquired and off-line processed to study the motional sidebands, and it is also sent to a lock-in amplifier and demodulated at $\Wpar/2$. The two quadrature outputs of the lock-in are simultaneously acquired and off-line processed. 

The third beam (pump beam),
orthogonally polarized with respect to the probe, is also sent to the cavity. Its field contains a main, cooling  tone at a frequency $\omega_{\mathrm{cool}}$ 
red detuned from the cavity resonance, 
%by roughly half a linewidth $(\Delta_{\mathrm{cool}}= -2 \pi \times 700 \mathrm{kHz} \simeq -\kappa/2)$, 
and a modulation tone at a frequency $\omega_{\mathrm{par}}$ blue shifted with respect to $\omega_{\mathrm{cool}}$. The two tones are obtained by driving the AOM on the pump beam with the sum of two radiofrequency signals. To realize the parametric drive of the oscillator, the frequency difference between the two tones is $\,(\omega_{\mathrm{par}} - \omega_{\mathrm{cool}}) = \Wpar = 2 \Omegam$. To obtain comparable oscillator spectra in the absence of coherent parametric drive, we further shift the modulation tone by $\sim 12$ kHz, i.e., by a quantity larger than the mechanical width, but much smaller than the cavity width. The opto-mechanical effect of the modulation tone is thus maintained almost constant, but the coherent effect of the two tones on the oscillator is avoided. During the measurements, the two values of the modulation tone frequency are alternated every 5s and the corresponding periods extracted from the acquired time series are analyzed separately. In this way, we can compare the two situations (with and without coherent parametric drive) keeping the same system conditions, avoiding the effect of possible long term drifts.     

All the radiofrequency sinusoidal signals used in the experiment, for driving the AOMs and as reference in the lock-in amplifier, are kept phase coherent. The system thus realizes a phase-sensitive heterodyne detection.

\begin{figure}
\begin{centering}
\includegraphics[scale=0.5]{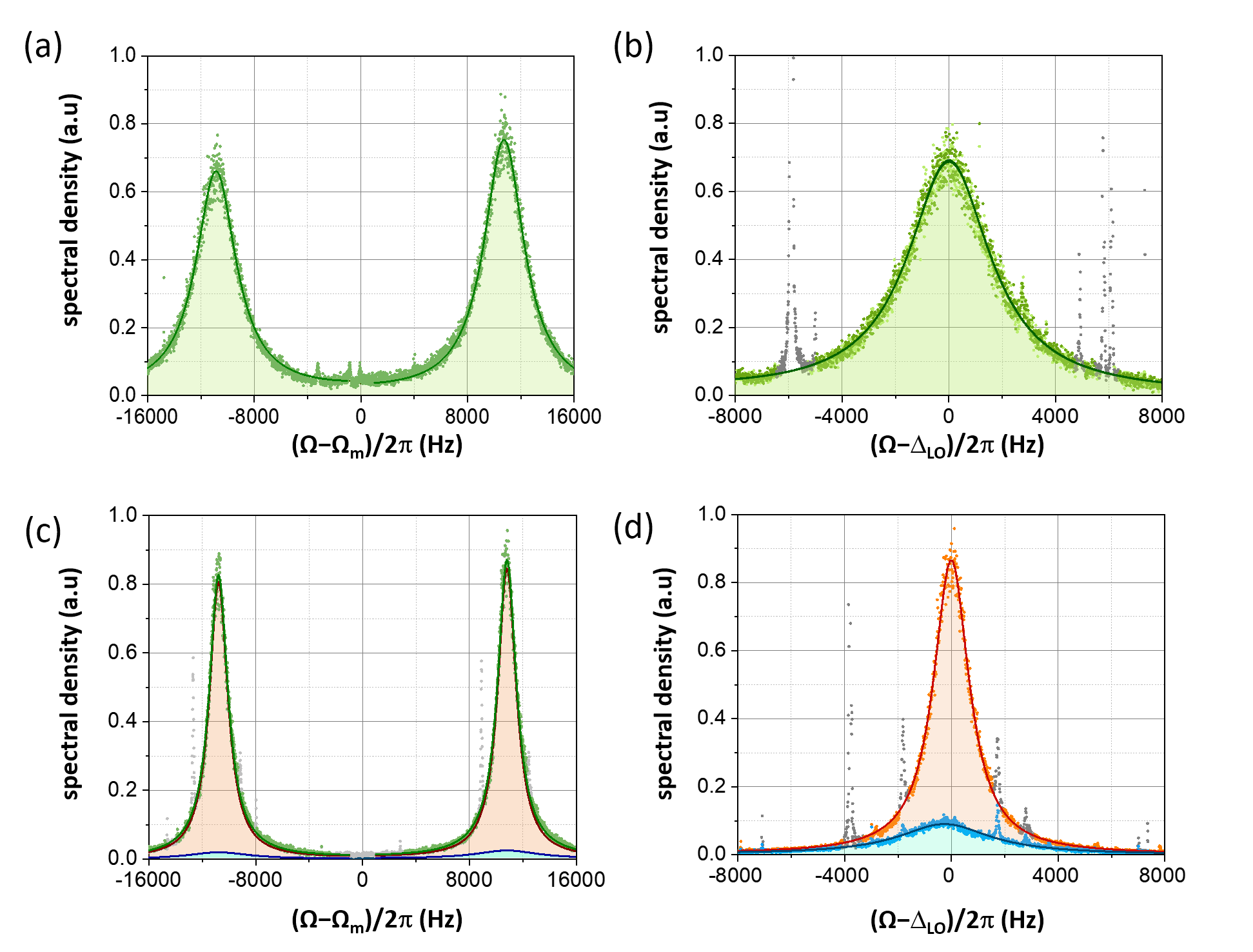}
%\par
\caption{(a),(c): Heterodyne spectra around the resonance frequency of the (0,2) membrane mode at $\Omegam/2\pi \simeq 530\,$kHz. The Stokes and anti-Stokes motional sidebands are visible around $\pm \Delta_{\mathrm{LO}}/2\pi = 11\,$kHz. In (a) the spectrum is acquired without resonant parametric drive, and is fitted by Lorentzian curves (solid line). In (c) (with parametric drive) the fitting function (dark green line) is the superposition of a broad and a narrow Lorentzian shape, whose contributions are shown with blue and red lines. (b), (d): Spectra of the fluctuations in two quadratures, obtained by phase-sensitive demodulation of the heterodyne signal at $\Wpar/2$. The oscillator quadratures $X$ and $Y$ originate spectral peaks around $\Delta_{\mathrm{LO}}/2\pi$. In (b), without parametric drive, the two spectra (shown respectively with dark and light green symbols) are not distinguishable, and one single fitting Lorentzian shape is shown with a solid line. In (d) (with parametric drive) the two spectra (shown respectively with orange and blue symbols) are fitted with different Lorentzian curves, shown respectively with red and blue solid lines. Few spurious electronic peaks, discarded from the data used for the fits, are shown with light gray symbols.} 
\label{fig_spettri}
\end{centering} 
\end{figure}

Typical heterodyne spectra are shown in Fig. (\ref{fig_spettri}a,c), together with the corresponding spectra of two orthogonal quadratures (Fig. \ref{fig_spettri}b,d). For the latter, the demodulation phase is chosen in order to produce maximally squeezed quadratures in case of resonant parametric drive. 

\begin{figure}
\begin{centering}
\includegraphics[scale=0.5]{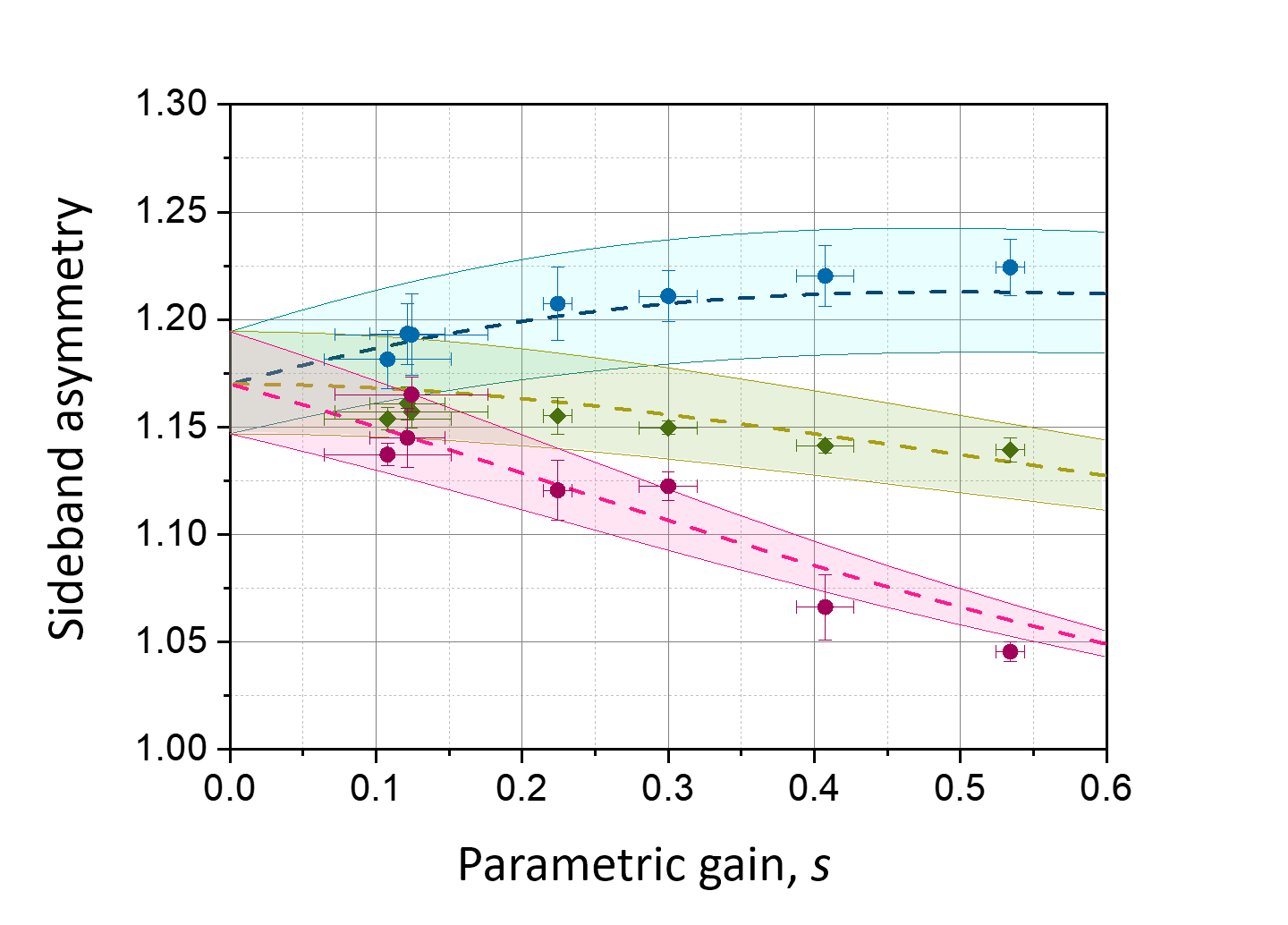}
%\par
\caption{Green symbols: sideband asymmetry $R$ in the absence of coherent parametric drive (i.e., with detuned modulation tone), for increasing power in the modulation tone. Ratio between the areas of the broad ($\Rplus$, shown with blue circles) and narrow ($\Rminus$, shown with red circles) Lorentzian contributions in the two motional sidebands, with coherent parametric drive. The values of the abscissa are the squeezing parameters $s$ extracted from the fitted widths $\Gplus = \Geff (1+s)$ and $\Gminus = \Geff (1-s)$. Dashed lines show the corresponding theoretical behavior, with shadowed areas given by the uncertainty in the system parameters (in particular, $5 \%$ in the cavity width and  0.5 K in the temperature). Error bars reflect the standard deviations in 5 consecutive independent measurements, each one lasting 100s. The total pump power is kept constant during all the measurements.}
\label{fig_RvsGain}
\end{centering} 
\end{figure}

The spectra shown in the upper panels (Fig. \ref{fig_spettri}a,b) are obtained without coherent parametric effect. The heterodyne spectrum consists of the two motional sidebands, separated by $2 \Delta_{\mathrm{LO}}$,  whose signal shapes are fitted by Lorentzian curves having the same width $\Geff$. The ratio $R$ of their areas, corrected for the residual probe detuning as described in Ref. \cite{Chowdhury2019}, allows to extract the oscillator occupation number $\nm$ according to $R = 1+1/\nm$. As shown with green symbols in Fig. (\ref{fig_RvsGain}), $R$ remains almost constant when varying the relative strength of the pump tones. A theoretical curve, based on independently measured parameters, shows indeed a weak dependence, due to the different cooling efficiency of the two pump tones. The agreement of this curve with the experimental data is good, and an extensive characterization of our system (reported in Ref. \cite{Chowdhury2019}) further confirms the reliability of the measurement of $\nm$.  

\begin{figure}
\begin{centering}
\includegraphics[scale=0.5]{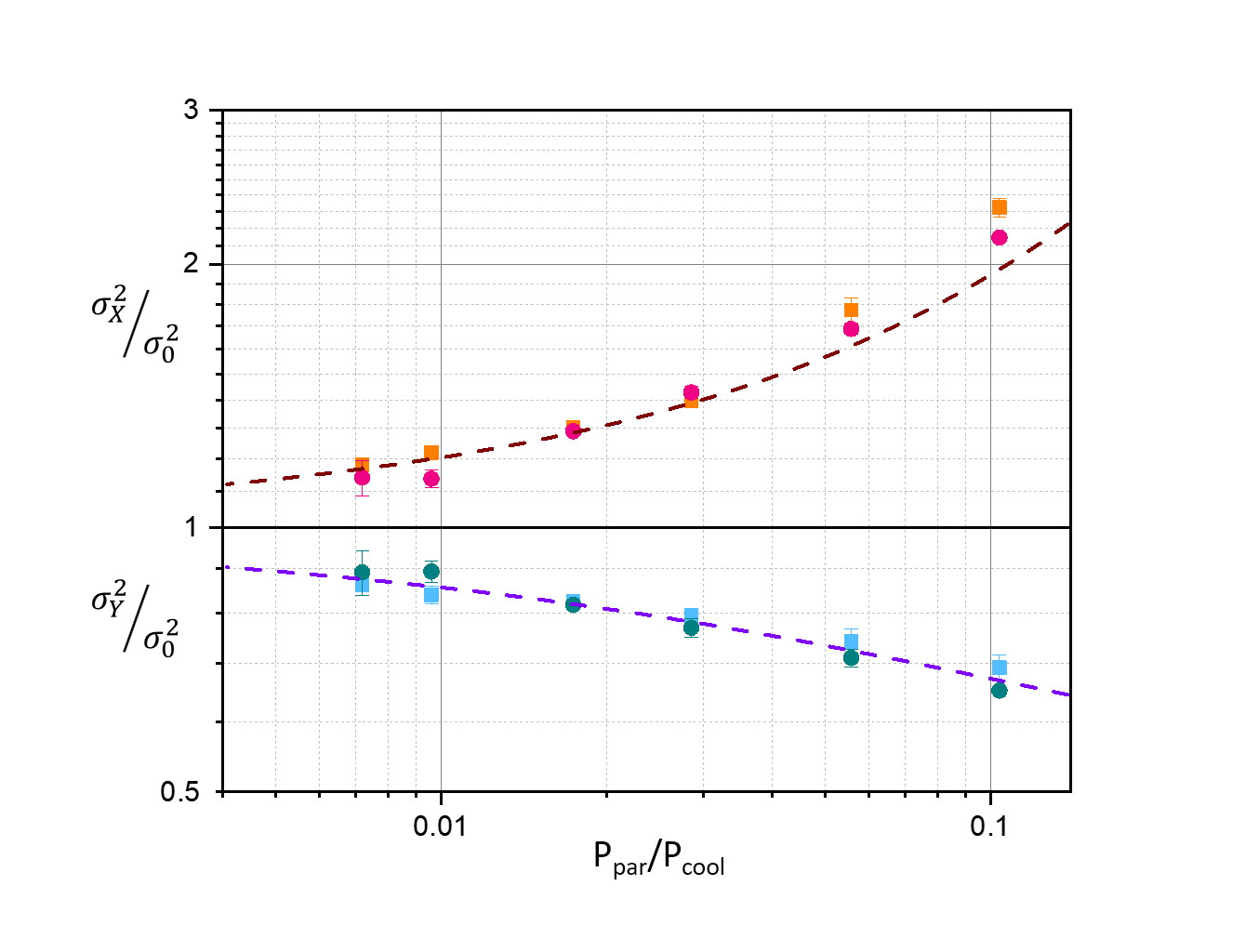}
%\par
\caption{Variance in the $X$ (orange squares) and $Y$ (cyan squares) quadratures, normalized to the value measured in the absence of coherent parametric drive, plotted as a function of the ratio between modulation and cooling tones, for constant total pump power. Dashed lines show the theoretical behavior, calculated with independently measured parameters (with no free fitting parameters). With magenta and blue circles we show the correspondent expected values, calculated respectively as $1/(1-s)$ and $1/(1+s)$, where the the parametric gain $s$ is extracted from the width of the broad and narrow Lorentzian contributions in the heterodyne spectra. Error bars reflect the standard deviations in 5 consecutive independent measurements, each one lasting 100s.}
\label{fig_gain}
\end{centering} 
\end{figure}

In the spectra of each quadrature (Fig. \ref{fig_spettri}b,d), the mechanical peak is visible at frequencies around $\Omega= \Delta_{\mathrm{LO}}$, and is originated by the superposition of the two motional sidebands. Each spectrum is fitted by the sum of two equal Lorentzian shapes centered at $\pm \Delta_{\mathrm{LO}}$:
\begin{equation}
S (\Omega) = \sigma_{0}^2 \left( \frac{\Gamma/2}{(\Omega-\Delta_{\mathrm{LO}})^2+(\Gamma/2)^2} +  \frac{\Gamma/2}{(\Omega+\Delta_{\mathrm{LO}})^2+(\Gamma/2)^2} \right)  
\label{Squad}  
\end{equation}
Without parametric effect (Fig. \ref{fig_spettri}b), the Lorentzian curves fitting the two quadratures turn out to be equal within the statistical uncertainty, and their width matches $\Geff$ extracted from the corresponding heterodyne spectra.
On the other hand, in case of resonant parametric drive the spectra of the two quadratures (Fig. \ref{fig_spettri}d) become respectively broader and narrower with respect to the previous case. The fitting function remains the same (\ref{Squad}), but with different areas ($\sigma_{X,Y}^2$) and widths ($\Gamma_{X,Y}$) for the two quadratures. The variances of the two quadratures, normalized to $\sigma_0^2$, are shown in Fig. (\ref{fig_gain}) as a function of the ratio between modulation and cooling tones, keeping a constant total pump power. Dashed lines show the expected behaviors, i.e., respectively $1/(1+s)$ and $1/(1-s)$, where the parametric gain $s$ is calculated using the measured pump tones ratio and detuning, and the cavity width $\kappa$. It can be expressed as $s = \Gamma_{\mathrm{par}}/\Geff$, with 
\begin{equation}
\Gamma_{\mathrm{par}} = \frac{4 g^2 \sqrt{\epsc (1-\epsc)} \, \Dpump}{\Dpump^2+\kappa^2/4}
\label{eq_Gpar}
\end{equation}
and
%\begin{eqnarray}
%\Geff = \Gm + g^2 \kappa & \left(\frac{\epsc}{\Dpump^2+\kappa^2/4}-\frac{\epsc}{(\Dpump-2\Omegam)^2+\kappa^2/4} \right.\nonumber\\
%& \left. +\frac{1-\epsc}{(\Dpump+2\Omegam)^2+\kappa^2/4}-\frac{1-\epsc}{\Dpump^2+\kappa^2/4} \right)
%\end{eqnarray} 
%\begin{equation}
%\Geff = \Gm + g^2 \kappa \left( \frac{\epsc}{\Dpump^2+\kappa^2/4}-\frac{\epsc}{(\Dpump-2\Omegam)^2+\kappa^2/4} 
%+ \frac{1-\epsc}{(\Dpump+2\Omegam)^2+\kappa^2/4}-\frac{1-\epsc}{\Dpump^2+\kappa^2/4} \right)
%\end{equation} 
\begin{equation}
\begin{split}
\Geff = \Gm + g^2 \kappa & \left(\frac{\epsc}{\Dpump^2+\kappa^2/4}-\frac{\epsc}{(\Dpump-2\Omegam)^2+\kappa^2/4} \right.\nonumber\\
& \left. +\frac{1-\epsc}{(\Dpump+2\Omegam)^2+\kappa^2/4}-\frac{1-\epsc}{\Dpump^2+\kappa^2/4} \right)
\end{split}
\label{eqGeff}
\end{equation} 
($g$ is the total optomechanical coupling strength of the pump beam, $\epsc$ is the ratio between cooling tone power and total pump power, $\Dpump$ is the mean detuning of the pump tones with respect to the cavity resonance, $\Gm$ is the mechanical width in the absence of optomechanical effects). We stress that in the expression of $s$, the optomechanical coupling $g$ disappears, eliminating the necessity of its, not obvious, evaluation. 
The shown theoretical lines are calculated with no free parameters, and we remark their excellent agreement with the experimental data.

The described imbalance between the fluctuations of the two quadratures is the characteristics of a squeezed state, but distinguishing a classical (thermal) state from a quantum state on this basis is not straightforward. The qualitative behavior is the same, therefore an accurate absolute calibration of the displacement variance and its comparison with that of the ground state 
%\viola{and, in general, a thorough comparison of the output spectra with a detailed model} 
are necessary. In the absence of parametric drive, the measured occupation number is here $\nm = 5.8$, therefore the variance in each quadrature is still $\sim 12$ times larger than in the ground state. At the maximum displayed squeezing, $\sigma^2_Y$ is reduced by a factor of 0.66, remaining well above that of the ground state. However, 
as we will show, the heterodyne spectrum already provides a direct quantum signature of the squeezed state. This is the most important result of our work. The spectral shape of each motional sideband departs from a simple Lorentzian peak, and it is indeed fitted by the sum of two Lorentzian curves with the same center, but different amplitudes and widths (Fig. \ref{fig_spettri}c). For the fitting procedure on the pair of motional sidebands we have used four independent Lorentzian amplitudes, and two widths written as $\Gplus = \Geff (1+s)$ and $\Gminus = \Geff (1-s)$, where $\Geff$ is fixed to the value derived from the corresponding spectra in the absence of resonant parametric drive. $\Gplus$ and $\Gminus$ agree, within the statistical uncertainty, respectively with $\Gamma_Y$ and $\Gamma_X$. Moreover, the parametric gain $s$ obtained from the fitted Lorentzian widths is in agreement with its estimate extracted from the variances of the two quadratures. This is displayed in Fig. (\ref{fig_gain}), where we show with circles the values of $(1+s)$ and $(1-s)$ given by the widths of the heterodyne spectra. We remark here the overall coherence between measurements of the variance in the quadratures (squares), measurements of the widths of the Lorentzian components in the heterodyne spectra (circles), and theoretical model (dashed lines).

The four areas of the Lorentzian components allow to calculate one sidebands asymmetry $\Rminus$ for the narrower Lorentzian peak, and one $\Rplus$ for the broader peak. These ratios are plotted in Fig. (\ref{fig_RvsGain}) as a function of the parametric gain $s$. The theoretical spectra of the Stokes and anti-Stokes sidebands are proportional respectively to
\begin{eqnarray}
S_{\bd \bd} & = & \frac{\Geff}{2}\left[\frac{1+\nm-s/2}{\Omega^2+\Gminus^2/4}+\frac{1+\nm+s/2}{\Omega^2+\Gplus^2/4}\right]    \label{eq_left}\\
S_{\bh \bh} & = & \frac{\Geff}{2}\left[\frac{\nm+s/2}{\Omega^2+\Gminus^2/4}+\frac{\nm-s/2}{\Omega^2+\Gplus^2/4}\right]   \label{eq_right}
\end{eqnarray}
The ratios between the areas of the broad and narrow Lorentzian components give the expressions (\ref{ratios1},\ref{ratios2}). The latter theoretical curves are also traced in Fig. (\ref{fig_RvsGain}), without free fitting parameters. The agreement with the experimental measurements is an indication that the overall understanding of the system is correct. Even with a moderately warm oscillator the analysis of the motional sidebands allows to extract and explore the quantum component of the motion of the mechanical oscillator in a squeezed state.
  
These results show that, as it happens for the thermal states of the oscillator, even for the squeezed state the transition between classical and quantum behavior is smooth, and a non-classical squeezing indicator is in principle measurable at any temperature. In other words, the quantum dynamics is present even in macroscopic oscillators dominated by thermal noise. It is however interesting to consider what should happen when the fluctuations in the squeezed quadrature are reduced below those of the ground state. This occurs for $(2\nm+1)/(1+s) < 1$, i.e., for $s > 2\nm$. From Eq. (\ref{eq_right}) we see that the broad Lorentzian contribution to the anti-Stokes sideband becomes negative (even if, of course, the overall spectral density remains positive for any $\Omega$). This is a threshold providing a clear indication that one is entering the \emph{ bona fide} quantum squeezing regime, without the necessity of absolute calibrations (we stress however that the increased sideband asymmetry is already a \emph{per se} quantum feature). The stationary parametric drive is stable for $s < 1$, therefore the above condition requires an initial occupation number $\nm < 0.5$. For thermal oscillators, several techniques have been conceived and demonstrated to overcome the parametric instability threshold in optomechanical systems, based on weak measurements and feedback \cite{Pontin_PRL2014,Sonar2018,Bowen2011,Bowen2013,Poot2014,Poot2015}. The extension of our analysis to this regime, as well as to the evolution of non-stationary squeezed states \cite{Pontin2016}, would provide additional insight to the quantum behavior of macroscopic mechanical systems, particularly useful for developing new protocols in the field of quantum sensing. 

Research performed within the Project QuaSeRT funded by the QuantERA ERA-NET
Cofund in Quantum Technologies implemented within the European Union' s
Horizon 2020 Programme. The research has been partially supported by INFN
(HUMOR project).

\end{document}